\documentclass[sigconf,nonacm]{acmart}

\acmConference[ICSE 2024]{46th International Conference on Software Engineering}{April 2024}{Lisbon, Portugal}

\usepackage{color}
\usepackage{amsmath}    
\usepackage{tabularray}
\usepackage{multirow}
\usepackage{natbib}
\usepackage{graphicx}
\usepackage{subfigure}
\usepackage{textcomp}
\usepackage{xcolor}
\usepackage{algpseudocode}
\usepackage{algorithm}

\theoremstyle{definition}

\usepackage{graphicx}
\usepackage{tabularx}
\usepackage{amsmath,xparse,mleftright}
\usepackage[export]{adjustbox}
\usepackage{balance}
\usepackage{tcolorbox}
\usepackage{algorithm,algpseudocode}
\usepackage{bbm}
\usepackage{dsfont}
\usepackage{url}

\usepackage{enumitem}
\usepackage{xcolor}
\usepackage{pifont}
\usepackage{xspace}

\usepackage{setspace}
\usepackage{url}

\newcommand{\eg}{\emph{e.g.},\xspace}

\newcommand{\etal}{\emph{et al.},\xspace}
\newcommand{\etc}{\emph{etc.}\xspace}

\newcommand{\name}{\textsc{Xpert}\xspace}
\newcommand{\company}{\textsc{Microsoft}\xspace}
\newcommand{\score}{\textsc{Xcore}\xspace}

\newcommand\boxwidth{8.5cm}
\newcommand\innerwidth{2mm}

\newcounter{insightC}

\newtcolorbox{mybox}[2][]{colbacktitle=red!10!white, colback=gray!10!white,coltitle=gray!70!black, title={#2},fonttitle=\bfseries,#1}

\acmDOI{}

\begin{document}

\title[\name: Empowering Incident Management with Query 
  Recommendations via Large Language Models]{\name: Empowering Incident Management \\ with Query 
  Recommendations via Large Language Models}

\settopmatter{authorsperrow=4} 

\author{Yuxuan Jiang}
\affiliation{%
\institution{University of Michigan Ann-Arbor}
\country{USA}
}\authornote{This work was completed during their internship at Microsoft Research Asia.}

\author{Chaoyun Zhang}
\affiliation{%
\institution{Microsoft}
\country{China}
}\authornote{Corresponding author.}

\author{Shilin He}
\affiliation{%
\institution{Microsoft}
\country{China}
}

\author{Zhihao Yang}
\affiliation{%
\institution{Peking University}
\country{China}
}\authornotemark[1]

\author{Minghua Ma}
\affiliation{%
\institution{Microsoft}
\country{China}
}

\author{Si Qin}
\affiliation{%
\institution{Microsoft}
\country{China}
}

\author{Yu Kang}
\affiliation{%
\institution{Microsoft}
\country{China}
}

\author{Yingnong Dang}
\affiliation{%
\institution{Microsoft}
\country{China}
}

\author{Saravan Rajmohan}
\affiliation{%
\institution{Microsoft}
\country{China}
}

\author{Qingwei Lin}
\affiliation{%
\institution{Microsoft}
\country{China}
}

\author{Dongmei Zhang}
\affiliation{%
\institution{Microsoft}
\country{China}
}


\acmConference[ICSE 2024]{46th International Conference on Software Engineering}{April 2024}{Lisbon, Portugal}




\begin{abstract}
Large-scale cloud systems play a pivotal role in modern IT infrastructure. However, incidents occurring within these systems can lead to service disruptions and adversely affect user experience. To swiftly resolve such incidents, on-call engineers depend on crafting domain-specific language (DSL) queries to analyze telemetry data. However, writing these queries can be challenging and time-consuming. This paper presents a thorough empirical study on the utilization of queries of KQL, a DSL employed for incident management in a large-scale cloud management system at \company. The findings obtained underscore the importance and viability of KQL queries recommendation to enhance incident management.


Building upon these valuable insights, we introduce \name, an end-to-end machine learning framework that automates KQL recommendation process. By leveraging historical incident data and large language models, \name generates customized KQL queries tailored to new incidents. Furthermore, \name incorporates a novel performance metric called \score, enabling a thorough evaluation of query quality from three comprehensive perspectives. We conduct extensive evaluations of \name, demonstrating its effectiveness in offline settings. Notably, we deploy \name in the real production environment of a large-scale incident management system in \company, validating its efficiency in supporting incident management. To the best of our knowledge, this paper represents the first empirical study of its kind, and \name stands as a pioneering DSL query recommendation framework designed for incident management.
\end{abstract}



\maketitle

\vspace*{-0.5em}
\section{Introduction}

Large-scale cloud systems, such as AWS, Azure, and Google Cloud, are indispensable pillars of modern IT infrastructure \cite{dillon2010cloud}. These platforms cater to a diverse range of online products and services, attracting a substantial global user base and generating significant revenue. In this context, ensuring the reliability of these cloud services becomes of utmost importance, as it directly impacts revenue generation and customer satisfaction \cite{yan2023aegis}. Despite considerable efforts invested in constructing robust systems, incidents continue to be a prevalent issue in cloud infrastructures \cite{grobauer2010towards}. Such incidents lead to service disruptions and have a detrimental impact on the overall user experience \cite{jin2023assess}.



In the event of an incident, on-call engineers (OCEs) diligently adhere to established procedures to swiftly identify the potential root cause and initiate prompt mitigation or resolution efforts \cite{chen2020towards, li2023conan}. To accomplish this objective, OCEs heavily rely on the analysis of telemetry data, encompassing vital runtime information such as logs \cite{zhang2021onion}, time series \cite{zhang2020microscope, zhao2023robust, zhang2019deep, zhang2021cloudlstm} and traces \cite{zeng2023traceark, wang2023root, chen2023imdiffusion}. These data hold a paramount significance in software systems and are consistently recorded and stored in databases throughout the service's operation. Similar to formulating standard database queries, OCEs manually compose domain-specific language (DSL) queries to extract the necessary telemetry data and thoroughly investigate the circumstances surrounding the incidents, enabling them to conduct effective triage and expedite the implementation of appropriate mitigation actions.

Writing appropriate queries for incident management is however a challenging task, as it necessitates considerable domain expertise to accurately select the appropriate databases, tables, and columns, and subsequently construct a query that incorporates various operations such as join, count, aggregation, \etc.
This process can be time-consuming and requires careful attention to detail. As a common practice, engineers often resort to finding suitable queries in existing troubleshooting guides (TSGs) to expedite the process \cite{shetty2022autotsg}. However, this approach relies heavily on the documentation ability of engineering teams, and OCEs may struggle to find the relevant queries if the TSGs are poorly organized or non-existent. Furthermore, queries can become highly complex, particularly when they span multiple databases. Even a minor mistake in the query can result in significant discrepancies in the retrieved results. Providing a useful query to OCEs can therefore significantly reduce their effort and expedite the incident mitigation process.

In this paper, we present a comprehensive empirical study on Kusto Query Language (KQL)
queries utilized for incident management in a world-wide cloud computing company \company. Our study delves into the frequency, complexity, and diversity aspects of these KQL queries. Key findings from our investigation reveal the following: \emph{(i)} the majority of incidents can be managed effectively with a small number of KQL queries; \emph{(ii)} Most of KQL queries used in incidents tend to be relatively simple in structure;  and \emph{(iii)} KQL queries exhibit long-tail pattern in templates and significant time variation. These insights underscore the necessity and practicality of automating KQL recommendation for OCEs to streamline the incident management.

Based on the insights gained from our empirical study, we introduce \name, an end-to-end framework designed to empower the incident management process by automatically recommending or generating KQL queries. Drawing from the abundant historical incidents and their corresponding KQL records in the past two years, \name provides customized KQL recommendations based on the specific context of new incidents. The framework efficiently extracts common patterns, such as tables and templates, from historical similar incidents, facilitating effective automation in \name. To address the limitations of traditional natural language processing (NLP) metrics \cite{papineni2002bleu} in evaluating domain specific queries, \name incorporates a novel performance metric called \score. This tailored metric allows for more comprehensive evaluation from three different perspectives, enhancing the overall quality assessment of the generated KQL queries.

To mitigate the challenges posed by costly pre-training, fine-tuning, and frequent updates of conventional NLP models, \name leverages the exceptional few-shot learning capabilities of Large Language Models (LLMs) to generate incident-specific KQL queries with only a few examples provided, without the need for parameter tuning. LLMs have demonstrated remarkable proficiency in parsing complex data \cite{le2023evaluation}, extracting essential information \cite{xiao2023enhancing}, and producing concise, insightful outputs in both natural language \cite{ding2023everything, mekala2022zerotop} and code \cite{vaithilingam2022expectation} domains. This makes them well-suited for the context of \name, where incident descriptions are often intricate and unstructured, posing challenges for traditional smaller language models. Moreover, the few-shot learning ability of LLMs in specific domains allows them to quickly adapt to novel and evolving incident types by leveraging historical data in an online fashion. This adaptability significantly enhances the quality of the generated KQL queries, rendering LLMs an ideal solution for this task.

We thoroughly evaluate \name, deploying it as a KQL recommendation framework, which serves as a pivotal component within the incident management system at \company. Experiments show the effectiveness of \name from both offline and online viewpoints. In summary, this paper presents the following contributions:

\begin{itemize}[leftmargin=*]
    \item A comprehensive empirical study on the utilization of KQLs in incident management, revealing interesting insights that inspire the KQL recommendation.
    \item Development of \name, an end-to-end KQL recommendation framework that leverages the few-shot learning capabilities of LLMs to empower automated KQL generation.
    \item Introduction of \score, an evaluation metric tailored to assessing the quality of generated KQL queries from various perspectives, addressing the limitations of traditional NLP metrics.
    \item Extensive offline evaluation of \name using a large-scale dataset from a real incident management system, showcasing its superior KQL quality compared to several strong baselines.
    \item Successful deployment of \name as a critical KQL recommendation framework within \name's incident management system, with pilot results demonstrating its exceptional performance.
\end{itemize}
To the best of our knowledge, this paper is the first to present an empirical study on the characteristics of DSL queries in incident management, and \name stands as the pioneering KQL recommendation framework specifically tailored for incident management.

\vspace*{-0.5em}
\section{Background}
This section provides an overview of incident management in the context of cloud computing, with a particular focus on the utilization of domain-specific language, KQL queries. 

\vspace*{-0.5em}
\subsection{Incident Management in Cloud}
The increasing popularity of cloud systems in recent years can be attributed to their inherent advantages, including scalability, accessibility, and cost-effectiveness \cite{chen2020towards, gupta2021review}. However, unplanned disruptions in cloud services, commonly referred to as incidents, remains a frequent phenomenon within cloud infrastructures \cite{ghosh2022fight, wang2021long}. To address incidents, OCEs typically rely on an incident management system, which involves various measures such as executing DSL queries, analyzing logs, and discussing with other engineers \cite{shetty2021neural, khang2022system}.

Taking the incident management system of \company as an example, when an incident is detected, a ticket is created in the system, with a title and summary provided by engineers manually or monitors automatically to describe the incident's context. 
During the incident management process, OCEs frequently compose and execute KQL queries on the service telemetry to comprehend the incident, identify the affected scope, and diagnose the underlying cause \cite{luo2014correlating, zhang2021onion}.
Typical telemetry data such as traces and logs related to the incidents may then be extracted and analyzed to aid in the diagnosis and mitigation processes \cite{ren2019time, chen2023imdiffusion, li2021fighting}.
These efforts yield valuable insights that greatly contribute to the triage process and subsequent mitigation actions \cite{ahmed2023recommending}.

\vspace*{-0.5em}
\subsection{KQL and KQL Queries}
A domain-specific language (DSL) \cite{mernik2005and} denotes a language that is specifically crafted for a distinct domain and customized to cater to specific tasks, industries, or areas of expertise. Illustrative examples of DSL queries encompass SQL for managing databases \cite{date1989guide}, GraphQL for querying graph databases~\cite{hartig2018semantics},  PromQL employed for querying metrics in Prometheus monitoring system \cite{sabharwal2020working},  Search Processing Language (SPL) in Splunk~\cite{splunk_spl}.

Kusto Query Language (KQL) is a DSL developed by \company, and it has been widely adopted within the organization. The language is designed to work with schema entities arranged in a hierarchy similar to SQL, comprising databases, tables, and columns. Behind the scenes, a big data analytics cloud service is optimized to handle KQL queries over various types of data, including structured, semi-structured, and unstructured data. For example, the services team regularly stores telemetry data such as traces, logs, and metrics in KQL databases, facilitating future queries for diagnostic and analytical purposes. By utilizing KQL, engineers can effectively explore the data, uncover patterns, identify anomalies, and perform other important tasks during the incident management process.




\begin{figure}[h]
\vspace*{-1.em}
\includegraphics[width=\columnwidth]{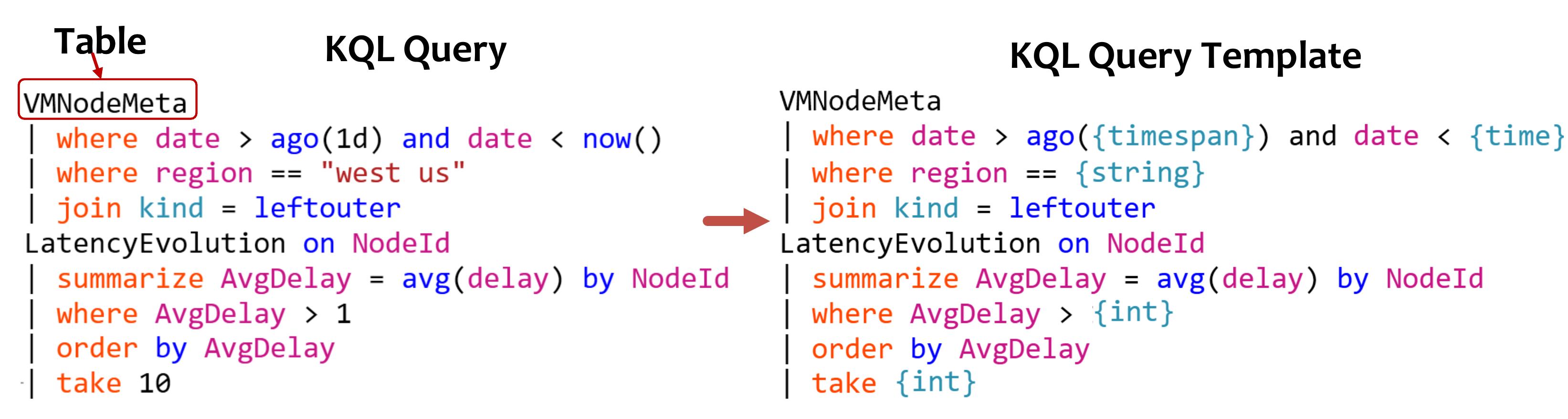}
\vspace*{-2.5em}
\caption{An example of a KQL query and its template.\label{fig:dsl}}
\vspace*{-1em}
\end{figure}

In Fig.~\ref{fig:dsl}, we present an example of a KQL query. A typical KQL query comprises a primary \emph{table name}, which signifies the main data source for the query. It may also include filter operators (\eg ``where''), selection operators (\eg ``take''), and join/union operators (\eg``join''). In essence, a query template can be derived from a KQL query by replacing the actual values in the query with placeholders. The resulting template is depicted on the right side of Fig.~\ref{fig:dsl}. These placeholders represent data types and query templates can be reused in different incident tickets. As a result, both the template and the full KQL query provide valuable information to OCEs for efficient incident management.

\vspace*{-0.5em}
\subsection{System Objective}
The primary objective of \name is to recommend KQL queries to OCEs within incident management systems, utilizing the available incident context and information. This recommendation process aims to facilitate various aspects of incident management, including triage \cite{pham2020deeptriage} and diagnosis \cite{he2022graph}, with the ultimate goal of reducing the time to mitigate (TTM) \cite{ghosh2022fight} for incidents. \name offers two hierarchical levels of KQL query recommendation:
\emph{(i)} \textbf{Template Recommendation}, which involves suggesting a query template (see Fig.~\ref{fig:dsl} right), that is the skeleton of queries for OCEs to fill in more concrete values.
\emph{(ii)} \textbf{Query Recommendation}, refers to predicting the full KQL query, with all values in the query filled. The recommended template and query are both submitted to the incident management system and presented to OCEs.

The rationale behind this hierarchical approach lies in the understanding that incident context may not always provide sufficient information for all values and fields in the full query. Some of these details may require domain-specific knowledge possessed by OCEs. By leaving certain fields as placeholders in the template recommendations, OCEs can complete the missing information based on their expertise. However, it is important to emphasize that template recommendations still provide valuable insights to OCEs, thereby enhancing the incident management process.

\vspace*{-0.5em}
\section{Empirical Study}
To better understand KQL in the context of incident management, we conducted a large-scale empirical study using incident tickets from the incident management system at \company. Our objective was to analyze the characteristics of KQL queries and gain insights to guide the design of \name. For this purpose, we collected 346,508 incident tickets that were categorized into four severity levels (0-3, from highest to lowest) from the top-30 services with the highest number of incidents, spanning from January 2021 to November 2022. These incidents represent over 60\% of total incidents and resulted in a dataset of 712,222 KQL queries. Note that only incident tickets that contain at least one KQL query are included. 
Specifically, we aimed to answer the following research questions (RQs):
\begin{itemize}[leftmargin=*]
    \item RQ1: How frequently are KQL queries used by OCEs?
    \item RQ2: How complex are the KQL queries used by OCEs?
    \item RQ3: How diverse are the KQL queries employed?
\end{itemize}
The following subsections present our findings to these RQs.

\vspace*{-0.5em}
\subsection{RQ1: Frequency of KQL Queries}
\begin{figure}[t]
\includegraphics[width=\columnwidth]{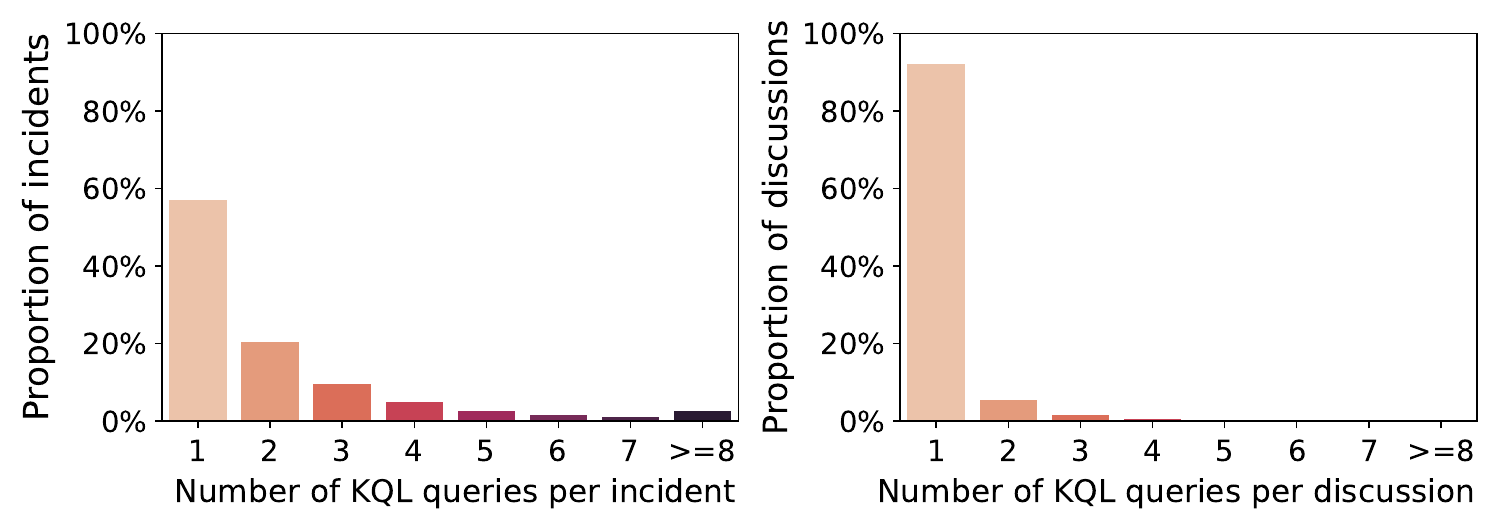}
\vspace*{-2.5em}
\caption{Distributions of KQL query number per incident/discussion.\label{fig:KQLnum}}
\vspace*{-1em}
\end{figure}


In the incident management platform, human-written KQL queries are typically posted in the ``discussion'' section of an incident ticket. The ticket may contain multiple discussions, and within each discussion, multiple KQL queries can be included. Fig.~\ref{fig:KQLnum} displays the distributions of the number of KQL queries per incident and per discussion across all incidents. Upon observation, it is evident that over half of the incidents have only one KQL query, and over 90\% of the discussions within an incident contain just one query. These findings indicate that a concise set of well-targeted queries is often sufficient to manage various incidents effectively. Given these findings, we set the primary objective of \name  to recommend the initial KQL query when a new incident is created, as this is sufficient to effectively address the majority of incident tickets.



\begin{table}[h]
    \centering
    \caption{Query reaction time statistics of incidents.\label{tab:time}}
    \vspace*{-1em}
    \begin{tabular}{cccccc}
    \hline
    Severity & Mean  & Median & P90 & P95 & P99 \\ \hline
    0        & 553   & 130    & 45  & 30  & 28  \\
    1        & 992   & 114    & 15  & 10  & 3   \\
    2        & 801   & 49     & 6   & 4   & 2   \\
    3        & 4,563 & 839    & 22  & 10  & 2   \\
    Overall  & 3,197 & 216    & 11  & 6   & 2   \\ \hline
    \end{tabular}
    \label{tab:Minutes}
\end{table}
Table~\ref{tab:time} presents the statistical analysis of the time duration between the occurrence of the first KQL query and the incident's creation time, referred to as ``query reaction time''.  This analysis is further categorized based on incident severity levels. Note that time unit in the table are normalized for anonymity. The query reaction time reflects the duration within which an OCE writes a KQL query in response to a reported incident.
Notably, for high-severity incidents (severity levels 0-2), the query reaction time is significantly shorter, indicating that OCEs respond much more promptly to these critical and impactful incidents. This observation aligns with our expectations, as high-severity incidents demand swift attention and action. The faster response times for high-severity incidents underscore the importance of timely intervention in critical situations, which can be supported by efficient KQL recommendations offered by \name.
Additionally, we observe that 95\% of incidents have a query reaction time greater than 6 units. This finding is valuable for informing the design of the time window of input that we include in \name, and its detailed design will be discussed in Sec.~\ref{tab:input}.



\begin{center}
\begin{tcolorbox}[colback=blue!5,
  colframe=gray!10,
  width=\boxwidth,
  arc=2mm, auto outer arc,
  boxrule=0.5pt,
  left=\innerwidth,
  right=\innerwidth,
  top=1pt,bottom=1pt
]
\textbf{Takeaways 1}: The number of queries required for effective incident resolution is small. The query reaction time for high-severity incidents is significantly shorter compared to lower-severity incidents. 
\end{tcolorbox}
\end{center}


\vspace*{-0.5em}
\subsection{RQ2: Complexity of KQL Queries}

\begin{figure}[t]
\includegraphics[width=\columnwidth]{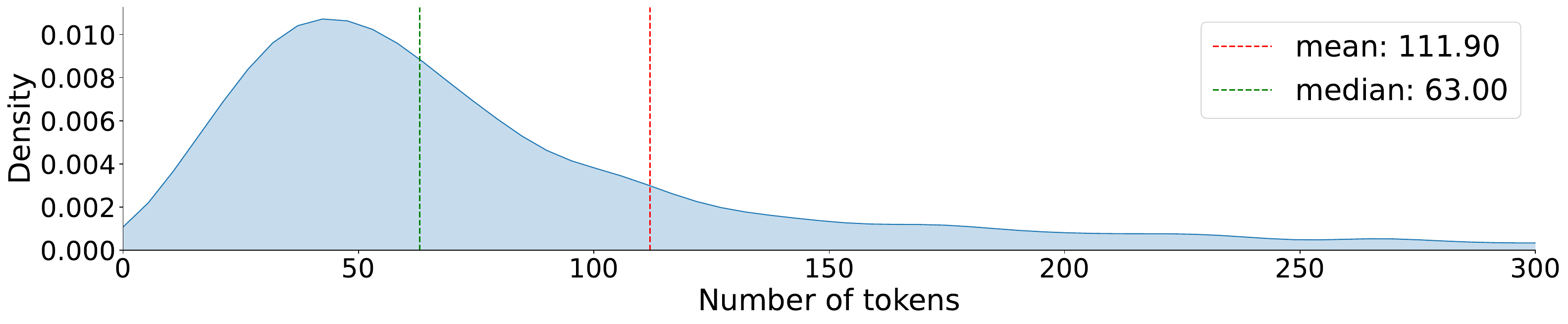}
\vspace*{-2.5em}
\caption{The token distribution of KQL queries in the dataset.\label{fig:token}}
\vspace*{-1em}
\end{figure}



We explore the complexity of KQL queries that OCEs write for incident management. The complexity of queries often signifies the level of filtering (\eg ``where'' operator) and the number of data sources involved, making it a valuable metric for gauging the comprehension and understanding of the incident.
To quantify the query complexity, we evaluate the number of tokens in the query, which provides a measure of the overall query length.

Fig.~\ref{fig:token} presents the distribution of tokens in KQL queries within the studied dataset. The analysis reveals that the token count follows a long-tail distribution, with the majority of queries being relatively short (less than 63 tokens). This aligns with our expectations, since during an incident, OCEs may have limited time for in-depth understanding, leading to the formulation of shorter queries. Consequently, this characteristic makes the task of query recommendation more feasible, as shorter queries are generally less diverse and easier to generate compared to more complex ones.


This findings reinforce the fundamental objective of \name. It highlights that in general, \name does not require the generation of overly complicated queries for most incidents, as the majority of incidents can be effectively managed with relatively concise and straightforward queries. This observation aligns with the design philosophy of \name, which aims to recommend KQL queries that are concise yet effective in addressing incident management tasks. 

\begin{center}
\begin{tcolorbox}[colback=blue!5,
  colframe=gray!10,
  width=\boxwidth,
  arc=2mm, auto outer arc,
  boxrule=0.5pt,
  left=\innerwidth,
  right=\innerwidth,
  top=1pt,bottom=1pt
]
\textbf{Takeaways 2}: The majority of incidents can be effectively managed using relatively concise and straightforward queries, rendering the query recommendation task more feasible.

\end{tcolorbox}
\end{center}




\vspace*{-0.5em}
\subsection{RQ3: Diversity of KQL queries\label{sec:diversity}}
\vspace*{-1em}


\begin{figure}[h]
\vspace*{-0.5em}
\includegraphics[width=\columnwidth]{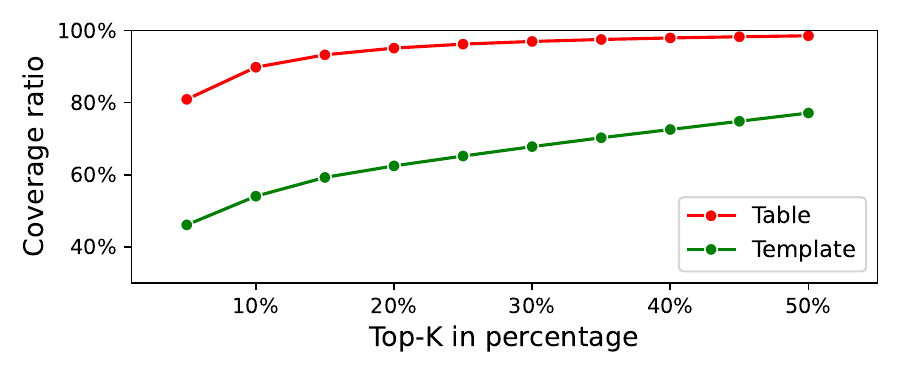}
\vspace*{-3em}
\caption{Query coverage ratio w.r.t. top-K tables/templates in percentage.\label{fig:coverage-figure-top30}}
\vspace*{-1em}
\end{figure}


Lastly, we turn our attention to investigating the diversity of KQL queries  for incident management. Fig.~\ref{fig:coverage-figure-top30} displays the coverage ratio with respect to the top-k percentage in terms of tables and templates. The results show that the top 5\% of tables used by most queries already cover 80.9\% of the queries, while the top 5\% of templates can cover 46.1\% of the queries. This long-tail pattern further supports the notion of low diversity in employed KQL queries, as common patterns are widely shared among them. This finding enhances the feasibility of the query recommendation goal, as these shared patterns can be effectively extracted by the LLMs. In addition, our analysis reveals an interesting finding regarding the sharing of query templates across different services. It is observed that query templates are rarely shared between services, with the percentage of shared templates being lower than 4.75\%. This low percentage indicates a high level of isolation between services, meaning that each service tends to have its own unique set of query templates specific to its incident management requirements. This finding holds important implications for the design of the hierarchical data retrieval approach in \name to limit the retrieval to incidents within the same service. We will provide details on this design in Sec.~\ref{sec:retrieval}.

\begin{figure}[h]
\includegraphics[width=\columnwidth]{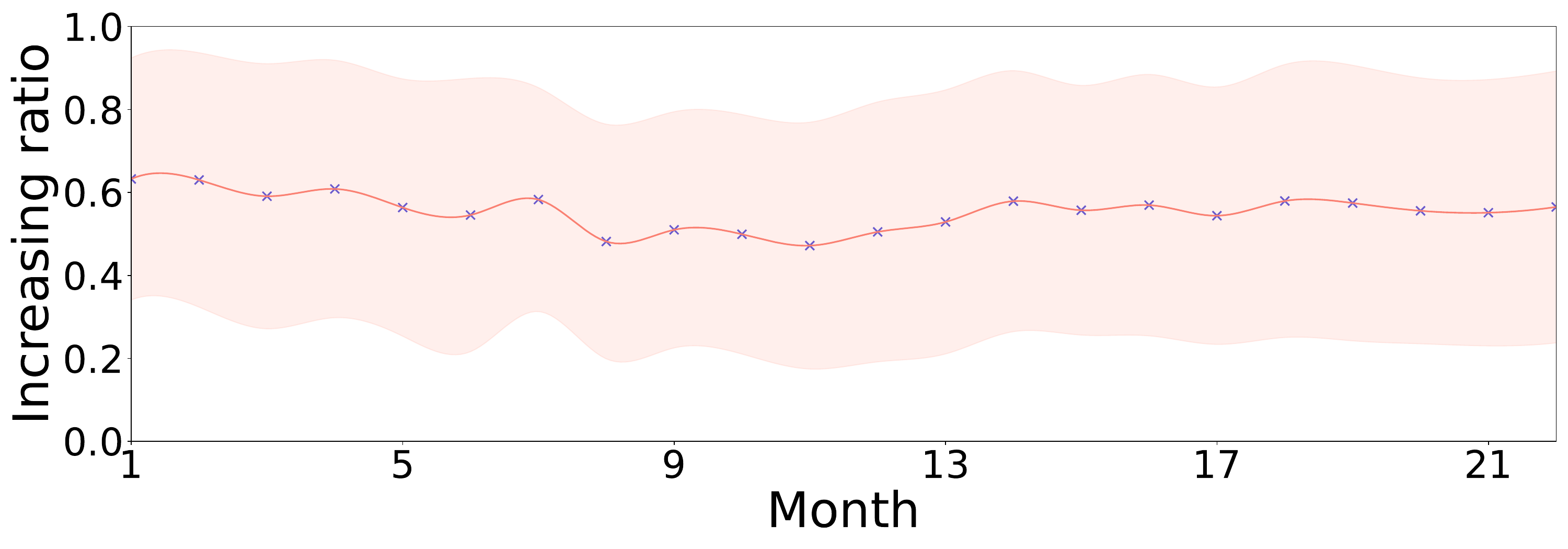}
\vspace*{-2.5em}
\caption{The  mean$\pm$std. of the monthly ratio of KQL queries covered by novel templates across all services.\label{fig:c_ratio}}
\vspace*{-1em}
\end{figure}

Finally, we present the mean$\pm$std. of the monthly ratio of KQL queries covered by novel templates across all services, as depicted in Fig.\ref{fig:c_ratio}. This metric measures the proportion of queries that are associated with previously unseen or unique templates within a given month, offering insights into the degree of change and data drift observed in KQL queries over time. We observe that KQL queries exhibit significant variation on a monthly basis, with an average of over 60\% of KQL queries utilizing different templates compared to the previous months, even within the same service. This finding highlights the dynamic and evolving nature of incidents, and underscores the need for adaptable and tailored KQL queries for incident management. Consequently, we have designed \name to accommodate such time-varying patterns in an online manner, as elaborated in Sec.~\ref{sec:online}.
\begin{center}
\begin{tcolorbox}[colback=blue!5,
  colframe=gray!10,
  width=\boxwidth,
  arc=2mm, auto outer arc,
  boxrule=0.5pt,
  left=\innerwidth,
  right=\innerwidth,
  top=1pt,bottom=1pt
]
\textbf{Takeaways 3}: \emph{(i)} KQL queries exhibit low diversity in terms of both tables and templates; \emph{(ii)} templates used in different service are rarely shared across each others; \emph{(iii)} KQL queries exhibit significant variation and data drift on a monthly basis.

\end{tcolorbox}
\end{center}

\noindent \textbf{Summary.} In summary, Takeaway 1 elucidates the primary objectives behind the design of our KQL query recommendation system \name. It underscores the importance of targeting the initial query and template of KQL queries for incidents, as these encompass a substantial portion of incident tickets. Furthermore, Takeaways 2 and 3 provide valuable insights into the relatively modest complexity and diversity observed in the KQL queries employed for incident management. These insights underscore the practicality of automating the recommendation of KQL queries. Subsequently, we introduce the architecture and design principles of \name, which are founded upon the discoveries derived from our empirical study.




\vspace*{-0.5em}
\section{The Design of \name}
We provide an overview of the architecture of \name in Section~\ref{sec:nutshell}, and delve into its  components in the following subsections.

\vspace*{-0.5em}
\subsection{\name in a Nutshell\label{sec:nutshell}}

\begin{figure*}[t]
\centering
\vspace*{-2.5em}
\includegraphics[width=\textwidth]{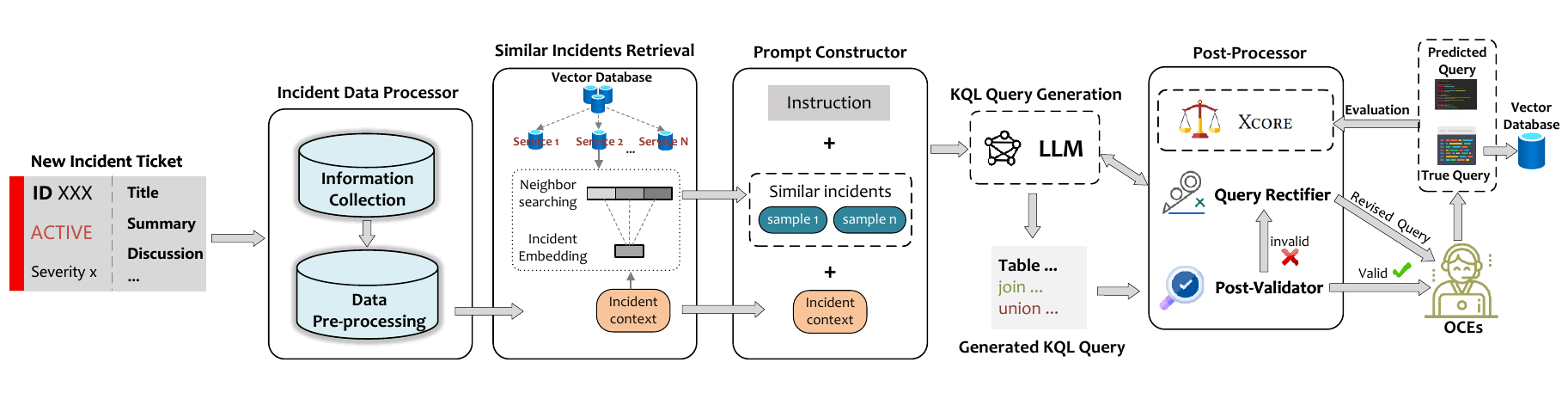}
\vspace*{-3.em}
\caption{The overall architecture of \name.\label{fig:framework}}
\vspace*{-1em}
\end{figure*}

Fig.~\ref{fig:framework} presents an overview of the \name framework. Upon receiving a new incident ticket in the incident management system, the data processor gathers relevant incident context and preprocesses it to a format compatible with the language model. Subsequently, an embedding model is employed to vectorize the incident context and conduct a search for similar historical incidents along with their corresponding KQL queries. By combining these retrieved samples with the target incident context in a prompt sequence, \name feeds this input into the LLM to generate an initial KQL template and query simultaneously.

The generated KQL template and query then undergoes a post-validation process to verify its adherence to correct grammar. Valid query is presented to the OCEs for recommendation, while invalid query undergoes necessary rectification to ensure its correctness before being forwarded to the OCEs. At last, the true queries crafted by OCE based on our recommendations are promptly added to the vector database to keep it up-to-date.  We also designed a dedicated \score to evaluate the quality of the recommended query, which we will elaborate in Sec.~\ref{sec:score}

\vspace*{-0.5em}
\subsection{Incident Data Processor}
The incident data processor gathers comprehensive information from the incident ticket and performs appropriate pre-processing to optimize the utilization of this data, as elaborated below.

\subsubsection{Information Collection\label{tab:input}} 
To equip the LLM with sufficient information for effective query recommendation, \name employs a comprehensive approach in collecting rich incident data from various resources within the incident management system \cite{chen2023empowering}. These resources encompass: \emph{(i)} \textbf{Metadata}, which entails fundamental incident details such as the creation time, the service which triggers the incident, and other essential information. \emph{(ii)} \textbf{Title} of the incident, which may be system-generated or written by an engineer. \emph{(iii)} \textbf{Summary} of the incident, serving as a high-level overview either generated by the monitoring system or written by an engineer. \emph{(iv)} \textbf{Discussion} pertaining to the incident, encompassing system logs related to the incident as well as discussions among the engineers. It is important to note that the discussions included in the incident context are confined to a time window of 5 time units from the creation time of the incident, ensuring the timeliness of the query recommendation.
The selection of this 5-unit time window is based on the findings presented in Table~\ref{tab:time}, which indicate that it covers at least 95\% of the incidents before the initial KQL query is posted. By incorporating these diverse contextual elements, \name maximizes the information available for the LLM, while upholding the timeliness of the query recommendation. 


\subsubsection{Data Pre-processing\label{sec:preprocess}} 
Upon collecting all information from the incident tickets, the data is concatenated into a text sequence. \name performs two pre-processing steps on the incident context: \emph{(i)} Repetitive information that appears multiple times in the context is removed. \emph{(ii)} If the incident context exceeds a certain token threshold, the sample is clipped to avoid over-length.  This is necessary as the input of the LLM is subject to token limitations. This pre-processing ensures improved information utilization in the data while adhering to the LLM's input constraints on token length.

\vspace*{-0.5em}
\subsection{KQL Query Recommendation}
Pretrained Language Models, such as the GPT series \cite{openai2023gpt4} and LLaMA \cite{touvron2023llama}, are typically trained on vast amounts of general information from publicly available domains or the Internet. However, their training corpus often lacks specialization in certain domains, such as internal incident management data. Fine-tuning a LLM with domain-specific data can be extremely costly \cite{rajbhandari2021zero}, and in some cases, it may be infeasible due to resource constraints or restrictions imposed by the model provider \cite{openai2023gpt4}.

To address this limitation, we explore the few-shot learning capability of LLMs \cite{brown2020language}, which allows us to demonstrate a few examples in the prompt sequence provided to the LLM for prediction. This approach, known as in-context learning (ICL), has shown to be effective in various domain-specific scenarios \cite{min2022rethinking}. By combining general knowledge from the LLM itself with context from specialized samples provided in the prompt sequence, ICL leverages both sources of information to make accurate predictions.

We therefore adopt a combination of LLM and ICL to generate KQL queries tailored for incident management. This approach utilizes the pre-processed incident context and similar historical incidents as demonstration and context for the LLM. The entire process encompasses three crucial steps,  namely \emph{(i)} similar incident retrieval, \emph{(ii)} prompt construction; and \emph{(iii)} DSL query generation. we provide  details of each step in the following subsections.

\subsubsection{Similar Incident Retrieval\label{sec:retrieval}}

The first crucial step in the ICL process is to retrieve incidents that are similar to the target incident, along with their corresponding queries. The ground truth queries are compiled when OCEs submit their initial query within the incident management system following the reception of an incident ticket.

This retrieval process involves utilizing an embedding model to vectorize historical incidents and the target incident, making them searchable using distance metrics. In this study, we utilize an embedding model \cite{muennighoff2022sgpt}, which encodes all incident contexts into 1,536-dimensional vectors. These vectors, along with their corresponding queries, are stored in a vector database called Faiss \cite{johnson2019billion}, enabling efficient data retrieval. 

When a new incident request is received, its context is embedded using the same embedding model. We then retrieve the top-$K$ similar incidents using the cosine similarity \cite{li2013distance} metric.
Since the retrieval process is performed exclusively  by considering only incidents falling within the same service as the target incident request, it narrows the retrieval to a more relevant range since data sources and templates are rarely shared across services, as suggested in Sec.~\ref{sec:diversity}. Furthermore, this exclusive retrieval significantly reduces retrieval costs. Once the top-$K$ similar incidents are obtained, we utilize them as context and construct the prompt sequence for the LLM, as detailed in the subsequent section.

\subsubsection{Prompt Construction and Query Generation}
\begin{figure}[t]
\includegraphics[width=\columnwidth]{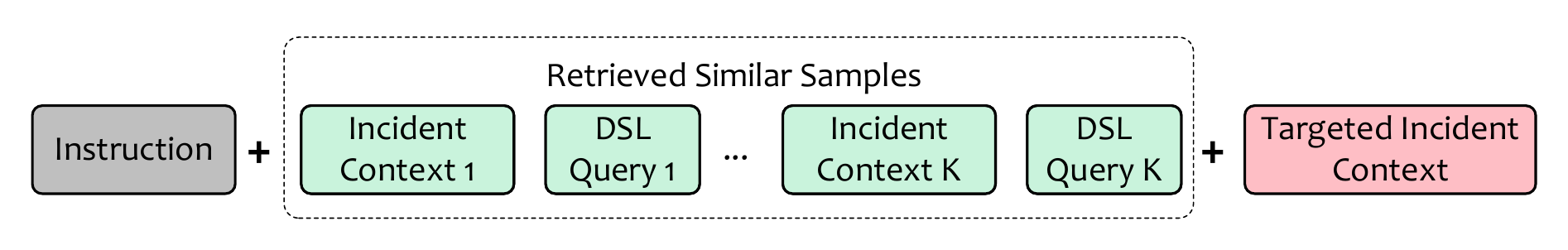}
\vspace*{-2.5em}
\caption{The structure of a prompt employed in \name.\label{fig:prompt}}
\end{figure}

Prompts as language sequence inputs to LLMs, serve as a means of interaction to accomplish specific tasks \cite{liu2023pre}. In the context of ICL, a typical prompt comprises three key elements, as illustrated in Fig.~\ref{fig:prompt}: \emph{(i)} An instruction, to inform the LLM about the goal of the task and the rules that apply to the task. \emph{(ii)} Retrieved similar incident samples from history, including the incident context and their corresponding true queries for incident management. \emph{(iii)} Targeted incident context, representing the incident context for which a KQL query needs to be recommended.

Note that the retrieved incident contexts undergo similar pre-processing steps as described in Sec.~\ref{sec:preprocess}. The number of retrieved samples included in the prompt depends on the total token length of the prompt. We adopt a greedy approach to add as many samples as possible to the prompt, maximizing the context and information provided to the LLM \cite{izacard2022few2}, while ensuring that the number of tokens in the prompt does not exceed the LLM's constraint (8k).
The constructed prompts are directly fed into the LLM to facilitate KQL query recommendation through the OpenAI API. The API returns a raw query that has been generated based on the prompt input.

\vspace*{-0.5em}
\subsection{Post-Processor}

To address the issue of potentially non-executable or grammatically incorrect KQL queries generated by LLMs, which may arise due to noise in retrieval data or mispredictions, we have integrated a post-processor into \name. The post-processor plays a crucial role in checking the validity of generated queries and rectifying any issues whenever possible. It comprises two key components:

\begin{itemize}[leftmargin=*]
    \item \textbf{Post-Validator:} This component performs a grammar and syntax check on the query using the intrinsic compiler abstract syntax tree (AST) \cite{neamtiu2005understanding}. By analyzing the data flow of the query, it determines if the query is executable. If the query fails this check, it is passed on to the post-rectifier for revision.
    \item \textbf{Post-Rectifier:} The post-rectifier carries out a two-step revision process to rectify invalid queries. In the first step, it cleans extraneous tokens from the query, such as spacing and tabs that might have been mistakenly generated. If the query still remains invalid, the post-rectifier proceeds to the second step, where we provide the LLM with the incident context, retrieved examples, the invalid query, error messages from the post-validator, and select usage handbook of the KQL. We then prompt the LLM to attempt fixing the query, resolving more complex cases that cannot be addressed by simple token removal.
\end{itemize}
This post-processing mechanism ensures that the KQL queries generated by \name are refined and enhanced to achieve executability and grammatical correctness, minimizing the need for manual intervention by OCEs. 
Once an OCE provides a ground-truth query for an incident, it is promptly added to the vector database. This ensures that the query becomes available for retrieval by future incidents, allowing the system to capture any potential data drift, as discussed in Sec.~\ref{sec:diversity}. 

\vspace*{-0.5em}
\section{Evaluation with \score\label{sec:score}}
Assessing the quality of generated KQL or other DSL queries poses a significant challenge, as effective metrics for such evaluations are lacking. Traditional NLP metrics, like BLEU \cite{post2018call} and METEOR \cite{banerjeelavie2005meteor}, primarily focus on lexical similarity and do not take into account code executability and execution accuracy. On the other hand, recent code metrics, such as CodeBLEU \cite{ren2020codebleu}, are designed to support various mainstream languages. However, these metrics need to be customized and tailored to the specific operators or DSL of a given language, rendering them less scalable. Moreover, evaluating the execution accuracy of KQL queries is a difficult task, further complicating the evaluation process, especially in the context of incident management. Execution results are sometimes not provided in the incident context and may not be easily reproducible. This limitation makes it challenging for other code metrics, such as Spider's SQL evaluation metric \cite{yu2019spider} and HumanEval \cite{liu2023code}, which rely on execution accuracy, to be effectively applied for KQL evaluation. Consequently, the overall evaluation of KQL quality remains a challenge.


\vspace*{-0.5em}
\subsection{Design of \score}


To address the limitations of current evaluation metrics, we introduce \score, an innovative assessment metric customized for evaluating the quality of generated queries, with a particular focus on KQL while also being adaptable to other DSLs. \score utilizes static code analysis \cite{louridas2006static} to parse KQL queries into distinct components using Abstract Syntax Trees (ASTs), and extract name reference nodes representing table columns. This parsing process enables us to perform a comprehensive evaluation of KQL queries from three key perspectives, namely \emph{(i)} Syntax and Semantic Check, \emph{(ii)} Sub-component Matching,  \emph{(iii)} Output-Schema Matching.  We detail each component in the following subsections. 

\vspace*{-0.5em}
\subsection{The Syntax and Semantic Check\label{sec:v}}
To evaluate the executability of a query, we utilize a two-step process. First, we employ the built-in syntax checker of KQL to detect any syntactical errors that may exist in the generated query. This step is crucial as it ensures that the query conforms to the language's grammar rules and structure. Next, the static analysis approach allows to approximately infer the semantic correctness of the query, by inferring the data types of these columns based on the semantics of the associated operations in the query. This analysis allows us to determine whether the query is semantically sound and executable.

\begin{figure}[t]
    \centering
    \includegraphics[width=\columnwidth]{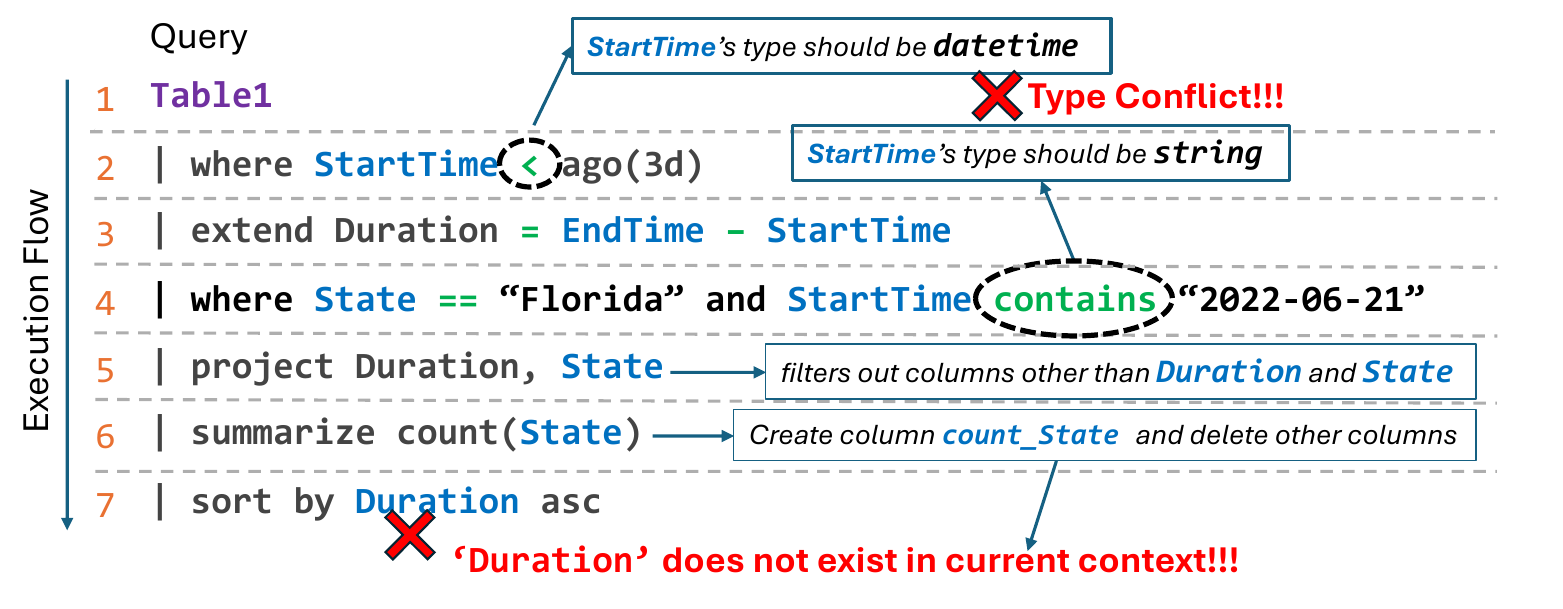}
    \vspace*{-2.5em}
    \caption{A semantic check example on an KQL query.\label{fig:score_exa}}
    \vspace*{-1.5em}
\end{figure}
Fig.~\ref{fig:score_exa} illustrates an example of the semantic check performed on an KQL query. In line 2 of the example, the type of \verb|StartTime| is inferred as \verb|datetime| based on the operand \verb|ago(3d)|. However, in line 4, the query calls a \verb|contains| operation, which only applies to \verb|string| data types. As a result, this operand becomes inexecutable due to a type mismatch. Additionally, in line 6, the \verb|summarize| operation drops the \verb|Duration| column, as it performs a ``groupby'' operation. However, in line 7, the \verb|Duration| column is called again, which makes the query invalid. Consequently, the semantic check assigns a zero score to this check, as the query contains invalid operations and is not executable. By conducting such semantic checks, we aim to estimate the validity of KQL queries, highlighting any issues that may arise during query execution.

It is important to note that this checker primarily assesses the query at the validity level, meaning that queries passing this checker are likely to be free of grammar mistakes. However, it is possible that queries passing this check may still encounter runtime errors during actual execution. Despite this limitation, the checker provides a valuable approximation of the query's validity and ensures that generated queries are less prone to syntax errors.


\vspace*{-0.5em}
\subsection{Sub-component Matching\label{sec:s}}

Leveraging the parsing results from the static analysis, we employ the sub-component matching technique to assess the lexical similarity between the generated queries and the ground truths. Following the design of Spider~\cite{yu2019spider}, we extract the operands of tabular components from both the ground truth and generated queries. For instance, for the code \verb|where colA > 20|, we extract the sets (\verb|colA|, \verb|where|, \verb|>|, \verb|20|) as representations of the operands of different tabular components. To measure the similarity, we compute the F1 score by comparing the sets of operands for both the ground truth and generated queries. The sub-component matching approach offers the advantage of considering operators that do not have strict order requirements \cite{xu2017sqlnet}. This means that changing the order of certain sub-components, such as filters, in a query may not necessarily alter the execution results. Therefore, the sub-component matching technique provides a more flexible and comprehensive evaluation of lexical similarity between  generated queries and  ground truths.


\vspace*{-0.5em}
\subsection{Output-Schema Matching\label{sec:o}}

The evaluation of execution accuracy includes a crucial step of assessing the output data schema, which comprises essential information such as column names, data source names, and output types (\eg tables or charts). This information plays a pivotal role in aiding engineers during incident troubleshooting, providing insights into aspects like impact start time and machine IDs. For instance, consider a scenario where valuable information pertains to the  \verb|response delay| field, which should be included in the query output. However, the final recommended query lacks this specific column, despite a high degree of similarity between the query code and the ground truth. The absence of such crucial information in the output diminishes the usefulness of the recommended query. Consequently, it becomes imperative to evaluate the correctness of the output-schema to address this issue.

To infer the output schema for both the generated and ground truth queries, we employ a similar data-flow analysis, as depicted in Fig. \ref{fig:score_exa}.
Once the output schema is obtained from the static code analysis, we parse it into a list of columns, a data source, and the return type. The final assessment of the output-schema matching is conducted in a two-fold manner: First, we use the F1 score to evaluate the similarity of output columns between the generated queries and the ground truth. Second, we employ binary accuracy to assess the correctness of the data source and the return types. This output-schema matching offers a different perspective on the information that should be returned to provide valuable insights, whose accuracy is particularly crucial in the context of incident management.

\subsection{Summarizing the Final \score}
Finally, we combine the results of the three evaluation components to encapsulate the overall quality of the generated query. We adopt a linear weighting scheme as follows:
\begin{equation}
\score = \alpha \cdot \mathcal{V} + \beta \cdot \mathcal{S} + \gamma \cdot \mathcal{O}.
\end{equation}

Here, $\mathcal{V}$ represents the binary validity score examined in Sec.\ref{sec:v}, $\mathcal{S}$ indicates the F1 score of the sub-component matching score described in Sec.\ref{sec:s}, and $\mathcal{O}$ represents the summarized output schema detailed in Sec.~\ref{sec:o}. $\alpha$,  $\beta$ and $\gamma$ are their corresponding weights, and they sum up to 1. In \score, we assign equal weights to $\mathcal{V}$, $\mathcal{S}$, and $\mathcal{O}$, as all three components hold comparable significance within the context of incident management. However,  these weights can be easily adjusted to prioritize specific aspects in different applications, if required.
 All three scores range from 0 to 1, making \score a continuous value ranging from 0 to 1 as well. A higher \score indicates a better quality of the generated query.



By integrating these three evaluation perspectives, \score provides a robust and comprehensive assessment of the quality of generated queries, effectively addressing the limitations of existing metrics and catering to the specific requirements of KQL. It is worth noting that while \score is specifically tailored to KQL, its adaptability to other DSL queries is straightforward. This can be achieved by simply substituting its AST and the built-in syntax checker. Hence, \score emerges as an adaptive and scalable metric applicable to various languages and scenarios beyond KQL.

\vspace*{-0.5em}
\section{Offline Evaluation}

\begin{table*}[t]
    \centering
    \caption{Performance evaluation of \name and other baselines in recommending KQL templates and full queries.\label{tab:performance}}
    \vspace*{-1em}
    \resizebox{1\textwidth}{!}{
\begin{tabular}{lcccccc|cccccc}
\hline
\multirow{2}{*}{\textbf{Model}} & \multicolumn{6}{c|}{Template}                                                                       & \multicolumn{6}{c}{Full Query}                                                                      \\ \cline{2-13} 
                                & BLEU           & METEOR         & \score         & TableAcc       & Identicality   & Validity       & BLEU           & METEOR         & \score         & TableAcc       & Identicality   & Validity       \\ \hline
Bart                            & 2.91           & 24.90           & 39.29          & 31.15          & 0.43           & 75.23          & 8.54           & 26.96          & 36.06          & 33.02          & 0.50            & 66.04          \\
T5                              & 60.90           & 60.84          & 38.17          & 31.02          & 10.52          & 49.98          & 30.29          & 46.65          & 31.10           & 23.94          & 4.64           & 33.89          \\
CodeT5                          & 70.48          & 69.64          & 59.35          & 48.21          & 18.70           & 77.60           & 64.44          & 64.88          & 57.42          & 52.39          & 14.59          & 73.49          \\
CodeT5+                         & 73.50           & 69.77          & 61.38          & 48.58          & 20.00          & 82.90         & 66.17          & 65.35          & 60.75          & 55.53          & 16.13          & 80.37          \\ \hline
\name (GPT-3.5)                 & 75.55          & 67.18          & 58.19          & 45.68          & 30.58          & 80.87          & 66.51          & 64.36          & 60.27          & 53.46          & 24.44          & 83.01          \\
\textbf{\name (GPT-4)}          & \textbf{76.89} & \textbf{71.61} & \textbf{62.40} & \textbf{48.81} & \textbf{35.46} & \textbf{83.27} & \textbf{70.45} & \textbf{67.98} & \textbf{64.2} & \textbf{57.56} & \textbf{29.18} & \textbf{86.34} \\ \hline
\end{tabular}
\vspace*{-1em}
}
\end{table*}

We conduct a comprehensive offline evaluation of \name, utilizing real incident data and KQL queries from the production environment of \company. We compare its performance against several baselines, to answer the following research questions (RQs):
\begin{itemize}[leftmargin=*]
    \item \textbf{RQ1:} How effective is \name in recommending KQL templates and queries?
    \item \textbf{RQ2:} How effective is the post-processor in correcting and refining the generated KQL queries?
    \item \textbf{RQ3:} How does \score outperform other NLP metrics?
\end{itemize}
We provide answers to these RQs in the subsequent subsections.

\vspace*{-0.5em}
\subsection{Experiment Setup}

\subsubsection{\textbf{Dataset}}
We use a large-scale dataset comprising incident context and corresponding KQL queries from the top-10 services with the most incident amount in \company. These services are the most representative and have the most significant impact. The data was partitioned into 197,666 instances for training and validation (for non-LLMs baselines) / retrieval (for LLMs), and 3,000 instances for testing. To prevent data leakage, incidents in the test data were strictly created after incidents in the training data.
Note that we perform one-shot offline evaluation and do not add the incidents to the vector database in an online manner during this evaluation.

\subsubsection{\textbf{Baselines}} 
We evaluate the performance of \name against several baselines, including smaller language models, namely: \emph{(i)} Bart \cite{lewis2020bart}, a popular sequence-to-sequence transformer model trained with denoising autoencoder fashion; \emph{(ii)} T5 \cite{raffel2020exploring}, an encoder-decoder transformer model pre-trained on a mixture of unsupervised and supervised tasks; \emph{(iii)} CodeT5 \cite{wang2021codet5}, an extension of T5 specifically designed for code-related tasks; and \emph{(iv)} CodeT5+ \cite{wang2023codet5+}, a more advanced version of CodeT5 pretrained with more diverse programming tasks and uses instruction tuning. All of the smaller language models are fine-tuned with training data. In this study, the terms ``T5'', ``CodeT5'' and ``CodeT5+'' refer to their base and 220M versions. For \name, we compare its GPT-3.5 and GPT-4 versions.

\subsubsection{\textbf{Evaluation Metrics}} We employ 6 evaluation metrics to comprehensively assess the quality of the generated KQL queries from various perspectives. In line with prior research \cite{gros2020code, ahmed2022multilingual, ahmed2023recommending}, we adopt two traditional NLP metrics, namely SacreBLEU \cite{post2018call} and METEOR  \cite{banerjeelavie2005meteor}, to evaluate the quality of the generated KQL queries. While these metrics have inherent limitations when assessing code quality, they can still offer valuable insights into certain aspects, such as the lexical  similarities. The proposed metric \score, addresses these limitations by evaluating the code from both syntax and semantic perspectives, making it a crucial measure for assessing the quality of KQL queries.

Additionally, we utilize TableAcc to quantify the accuracy of the predicted table names in the queries, which provides a partial data source assessment. Furthermore, Identicality offers a more stringent evaluation of the recommended queries, measuring the proportion of recommended queries that are identical to the ground truth. To ensure the executability of the generated queries, we also employ the Validity metric, evaluating the percentage of generated KQL queries that comply with the KQL grammar. This is a critical aspect to guarantee the executability of the generated queries.

\subsubsection{\textbf{Implementation}} 
The smaller language models were implemented using \texttt{Pytorch} \cite{paszke2019pytorch} and fine-tuned on 2 NVIDIA A100 GPUs. \name interacts with GPT Ada Embedding, GPT-3.5 and GPT-4 through the \texttt{Python} API provided by OpenAI. The entire implementation of \name consists of 1,929 lines of \texttt{C\#} code and 2,694 lines of \texttt{Python} code.


\vspace*{-0.5em}
\subsection{KQLs Recommendation Performance (RQ1)}
Table~\ref{tab:performance} presents the performance evaluation of \name in recommending KQL templates and full queries, compared to various baseline models. Notably, \name equipped with GPT-4 demonstrates superior performance across all metrics for both template and query recommendation. This outcome suggests that \name represents the state-of-the-art approach from various perspectives, encompassing improved lexical similarity metrics (BLEU and METEOR), enhanced syntax correctness and semantic analogy assessments (\score), higher table accuracy (TableAcc), and superior exact matching with the ground truth (Identicality) and executability (Validity). Particularly, its advantage is most pronounced in the Identicality dimension, where both GPT-3.5 and GPT-4 versions of \name significantly outperform other smaller language models. This aspect bears particular significance in the incident management scenario, as refining a query can be a laborious task for OCEs. A high Identicality rate in \name recommendations renders the process more user-friendly and accessible, particularly for less experienced OCEs.

CodeT5+ emerges as a stronger baseline compared to other baselines, exhibiting comparable performance with \name (GPT-3.5) in several metrics. However, note that CodeT5+ requires a costly fine-tuning process, rendering it difficult and expensive to adapt to evolving incident types in an online manner. In contrast, \name offers a more effective, simpler, and cost-efficient solution by effortlessly adding a new incident and its ground truth query to the vector database upon arrival, thereby capturing online data drift, as will be demonstrated in Sec.~\ref{sec:online}.

Moreover, as templates of KQLs are less diverse compared to full queries, \name achieves greater identicality in template recommendation compared to query recommendation. This can be particularly helpful in situations where incident context is incomplete. In such cases, OCEs can refine KQL queries based on the predicted templates, saving the query writing time. Overall, \name delivers remarkable performance in every dimension, especially when empowered by GPT-4. These inspiring results establish \name as an effective solution for recommending KQL queries, thereby significantly facilitating the job of OCEs during incidents.


Lastly, it is important to highlight the impressive few-shot learning capability of LLMs observed during our experiments. The average and median number of similar samples retrieved for the prompt sequence provided to \name, with a maximum of 8k tokens, is 7.41 and 6, respectively. These values are surprisingly small, indicating that LLMs can achieve remarkable performance with very few examples for demonstration. In fact, LLMs outperform other smaller models that are trained with approximately 200 thousand samples in all evaluated dimensions. This outstanding few-shot learning ability not only reduces the effort of pre-training and fine-tuning but also positions LLMs as an ideal solution for real productions.

\vspace*{-0.5em}
\subsection{Post-proceessor Effectiveness (RQ2)}
\begin{table}[t]
    \centering
    \caption{Comparison before and after the post-processing.\label{tab:post_processor}}
    \vspace*{-1em}
\begin{tabular}{lcc|cc}
\hline
\multirow{2}{*}{Metric} & \multicolumn{2}{c|}{Template} & \multicolumn{2}{c}{Full Query} \\ \cline{2-5} 
                        & Before    & \textbf{After}    & Before     & \textbf{After}    \\ \hline
BLEU                    & 34.95     & \textbf{36.61}    & \textbf{30.52}      & 27.88 \\
METEOR                  & 42.24     & \textbf{46.30}    & \textbf{47.77}      & 46.14    \\
\score                  & 11.63     & \textbf{35.99}    & 11.19      & \textbf{35.93}    \\
TableAcc                &22.54      & \textbf{38.03}    & 17.39      & \textbf{31.68}    \\
Identicality            & 0.00      & \textbf{7.04}     & 0.00       & \textbf{10.56}    \\
Validity                & 0.00      & \textbf{50.70}    & 0.00       & \textbf{56.52}    \\ \hline
\end{tabular}
\end{table}

We now shift our attention to evaluating the effectiveness of the post-processor in improving the query generation quality. In the raw predictions of \name (GPT-4), a total of 71 templates and 161 queries are found to be invalid. After applying the post-processor, 36 (50.70\%) templated and 91 (56.52\%) queries are successfully fixed. Taking a closer look at the impact of the post-processor on these invalid cases as shown in Table~\ref{tab:post_processor}, we observe that it significantly improves the quality of the generated queries across various metrics. The \score of generated templates and queries improved by 24.36 and 24.74 respectively, showcasing its efficacy in enhancing the overall query quality. Moreover, the identicality metric saw an improvement of 7.04\% and 10.56\% for templates and queries respectively, further highlighting the necessity and effectiveness of the post-processor. These results demonstrate that the post-processor is an indispensable component of \name, as it plays a vital role in refining and enhancing the generated queries.

\vspace*{-0.5em}
\subsection{\score Evaluation (RQ3)}
\begin{figure}[t]
    \centering
    \includegraphics[width=\columnwidth]{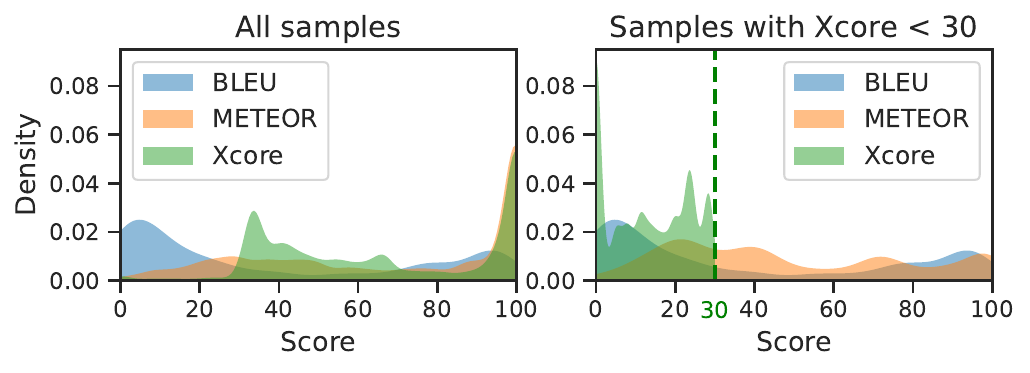}
    \vspace*{-2.5em}
    \caption{Distributions of BLEU, METEOR and \name in the test set (left), and samples with \score below 30 (right). \label{fig:score_cmp}}
\end{figure}

Finally, we present a comparison of \score with other NLP metrics to emphasize the effectiveness of our proposed design. In Fig.~\ref{fig:score_cmp}, we depict the distributions of BLEU, METEOR, and \score on all samples in the test set (left), as well as on selected samples where \score is below 30 (right). Upon examination, we observe distinct differences in the distributions of \score compared to the other two metrics. While \name and METEOR show significant overlap in the high score region (over 90), they exhibit substantial differences in the low score region (<30). Upon closer examination of the samples where \score falls below 30 in the right subplot, we observe that although \score values are low, both BLEU and METEOR metrics are evenly distributed across their entire range from 0 to 100. This observation prompts us to investigate the discrepancy among these metrics in these particular samples to determine which metric provides a more reasonable evaluation.

\begin{figure}[t]
    \centering
    \includegraphics[width=\columnwidth]{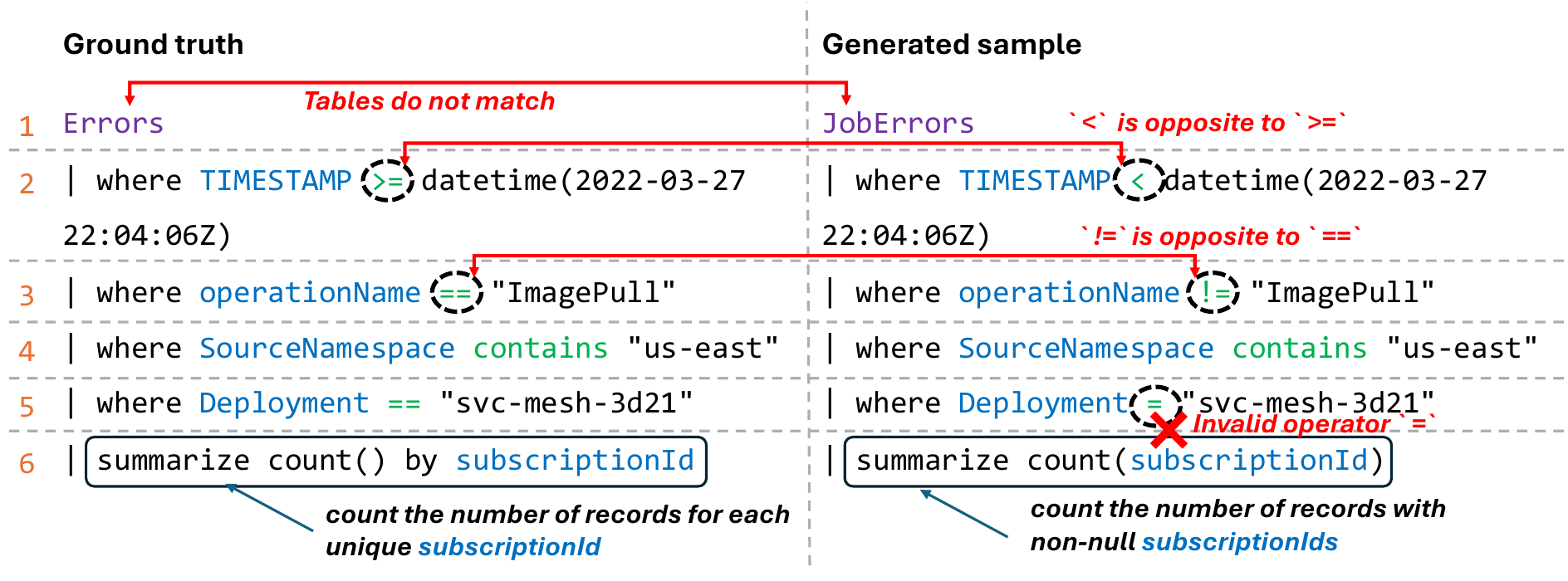}
    \vspace*{-2.5em}
    \caption{A representative generated sample and its corresponding ground truth. \label{fig:score_cmp_case}}
    \vspace*{-1.em}
\end{figure}

To this end, we present a representative generated sample and its corresponding ground truth in Fig.~\ref{fig:score_cmp_case}. The generated sample achieves satisfactory scores on NLP metrics (75.67 for BLEU and 87.02 for METEOR), as there is a high lexical similarity between the two queries. However, it only achieves a score of 3.54 on \score, which is significantly lower. Upon closer examination, we observe the following issues: \emph{(i)} the generated sample fails the validity check (Sec. \ref{sec:v}) due to the incorrect usage of \verb|=| in the \verb|where| operator (line 5); \emph{(ii)} various operations are inverted in the generated sample (lines 2 and 3), resulting in substantial sub-component mismatches (Sec. \ref{sec:s}); and \emph{(iii)} examination of the output schema (Sec. \ref{sec:o}) uncovers discrepancies in the source tables and output columns in the two queries (lines 1 and 6). These discrepancies result in low scores of $\mathcal{V}$, $\mathcal{S}$, and $\mathcal{O}$, leading to the low overall \score.

This discrepancy between \score and other NLP metrics highlights the effectiveness and reliability of \score for evaluating the quality of KQL queries. While the generated sample may appear plausible and lexically similar based on BLEU and METEOR, closer examination reveals its invalidity and inaccuracies, making \score a more robust and meaningful metric in this context.

\vspace*{-0.5em}
\section{Production Impact\label{sec:online}}
\begin{table}[t]
    \centering
    \caption{Performance comparison of \name in productions.\label{tab:online}}
    \vspace*{-1em}
    \resizebox{1\columnwidth}{!}{
\begin{tabular}{lcc|cc}
\hline
\multirow{2}{*}{\textbf{Model}} & \multicolumn{2}{c|}{Template}                                                                       & \multicolumn{2}{c}{Full Query}                                                                      \\ \cline{2-5} 
        & \score                & Identicality                 & \score         & Identicality      \\ 
        \hline
CodeT5+ (Offline)     & 58.5             &  10.04                         &  64.94             &  2.62  \\
\name (Offline)           & 55.63            & 14.19                  & 66.55       & 11.14           \\
\hline
CodeT5+ (Online)      & \textbf{58.79}           & 10.04          & 65.43        & 2.62            \\
\textbf{\name (Online)}   & 58.74 & \textbf{18.12} 
  & \textbf{72.96} & \textbf{17.69} \\ 
\hline
\end{tabular}
}
\vspace*{-1.5em}
\end{table}

The proposed framework \name, has been seamlessly integrated as a crucial component within the incident management system of \company. A pilot study was conducted, wherein \name was deployed for approximately one month on incidents in a representative service. Whenever an incident ticket is received, \name is triggered to recommend the first XQL query for incident management, following the workflow depicted in Fig.~\ref{fig:framework}. Unlike the offline evaluation, each time a new incident ticket and its corresponding ground-truth query written by OCEs are recorded, we promptly add this instance to the vector database. This functionality enables subsequent requests to retrieve this recorded sample, ensuring that \name can continuously adapt to new types of incidents without requiring adjustments to other models or settings. We refer to this framework equipped with GPT-4 deployed in the production environment, as ``Online \name''. To compare the effectiveness of \name in the real production environment, we conducted fine-tuning on CodeT5+ weekly (referred to as ``Online CodeT5+'') and used its latest version for predictions. For both \name and CodeT5+, we compare their performances without vector database updates and fine-tuning, denoting them as ``Offline'' versions.

The piloting results are presented in Table~\ref{tab:online}. The online experiments yield inspiring outcomes, revealing the superiority of \name over CodeT5+ in various aspects. Firstly, both online and offline versions of \name significantly outperform CodeT5+ in terms of \score and Identicality in XQL query recommendation, affirming \name's consistent advantage in over smaller language models. Secondly, the online version of \name demonstrates superior performance compared to its offline counterpart. This suggests that \name effectively captures the significant time variation of incidents through effortless vector database updates, enabling it to adapt more flexibly to the ever-evolving real production environment. Furthermore, the average online response time for \name is approximately 5 seconds, demonstrating its efficiency in meeting the online responsiveness requirement. All these results further underscore \name's potential as a more robust, efficient and adaptive solution for XQL recommendation in practical scenarios.

\vspace*{-0.5em}
\section{Discussion}

\subsection{Case Study}

\subsubsection{\textbf{Successful Case}} 
\begin{figure}[t]
    \centering
    \includegraphics[width=\columnwidth]{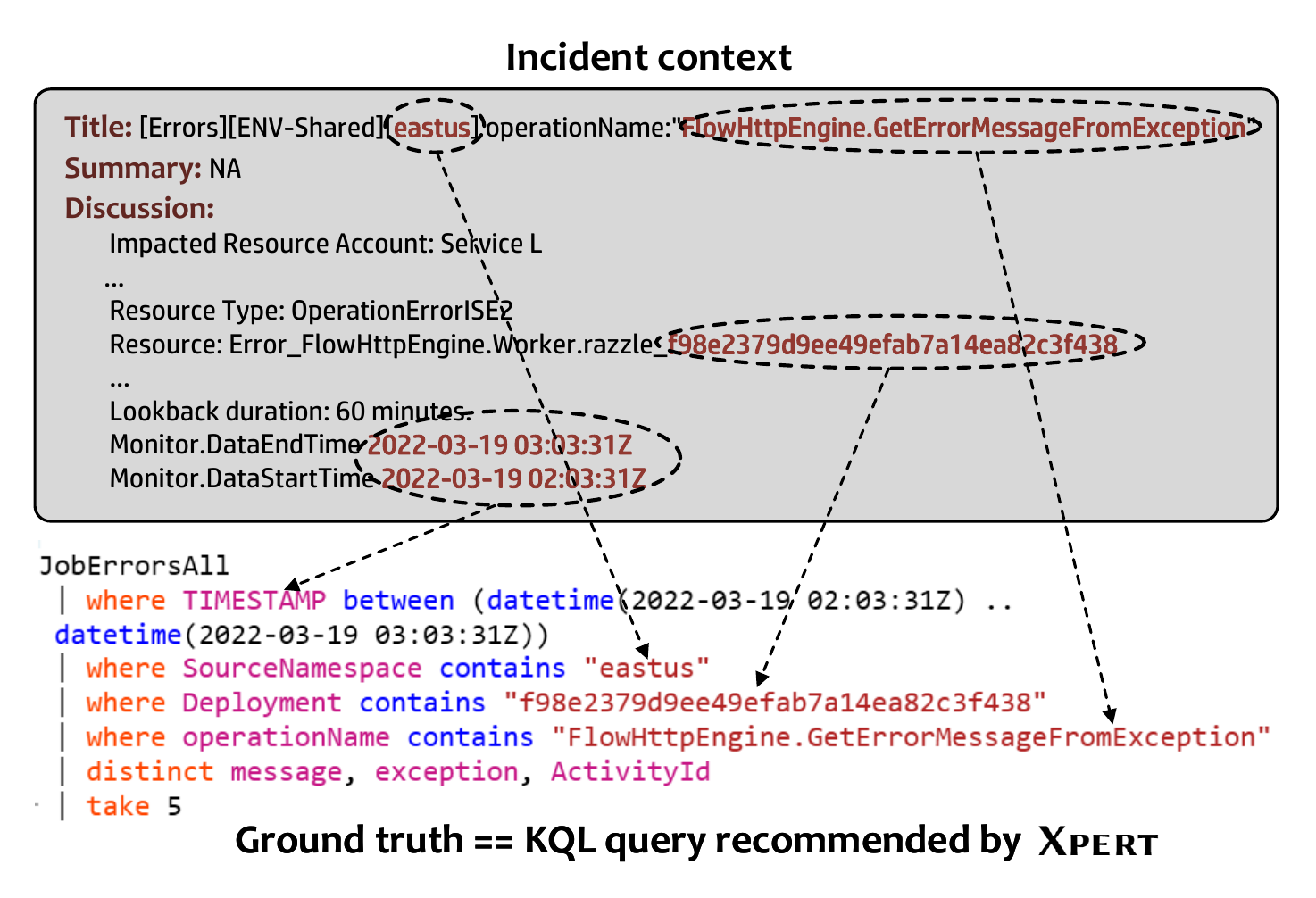}
    \vspace*{-2.5em}
    \caption{A successful case that \name recommends an KQL query identical to the ground truth. \label{fig:goodcase}}
    \vspace*{-1.5em}
\end{figure}

In Fig.~\ref{fig:goodcase}, we present a successful case where \name recommends an KQL query that perfectly matches the ground truth written by an OCE. In this case, \name efficiently extracts relevant information from the incident title and discussion, including details such as time, region, and resource ID. It skillfully places this extracted information in the appropriate positions within the query to construct the operands. By leveraging this context, \name successfully generates a query that is not only syntactically and semantically correct but also identical to the query written by the OCE. This outcome demonstrates the superior information extraction capability of LLMs, enabling them to distill valuable insights from complex and unstructured data. Consequently, LLMs are well-suited for the KQL query recommendation task and can provide significant assistance to OCEs.
\vspace*{-0.5em}
\subsubsection{\textbf{Failed Case}} 

\begin{figure}[t]
    \centering
    \includegraphics[width=\columnwidth]{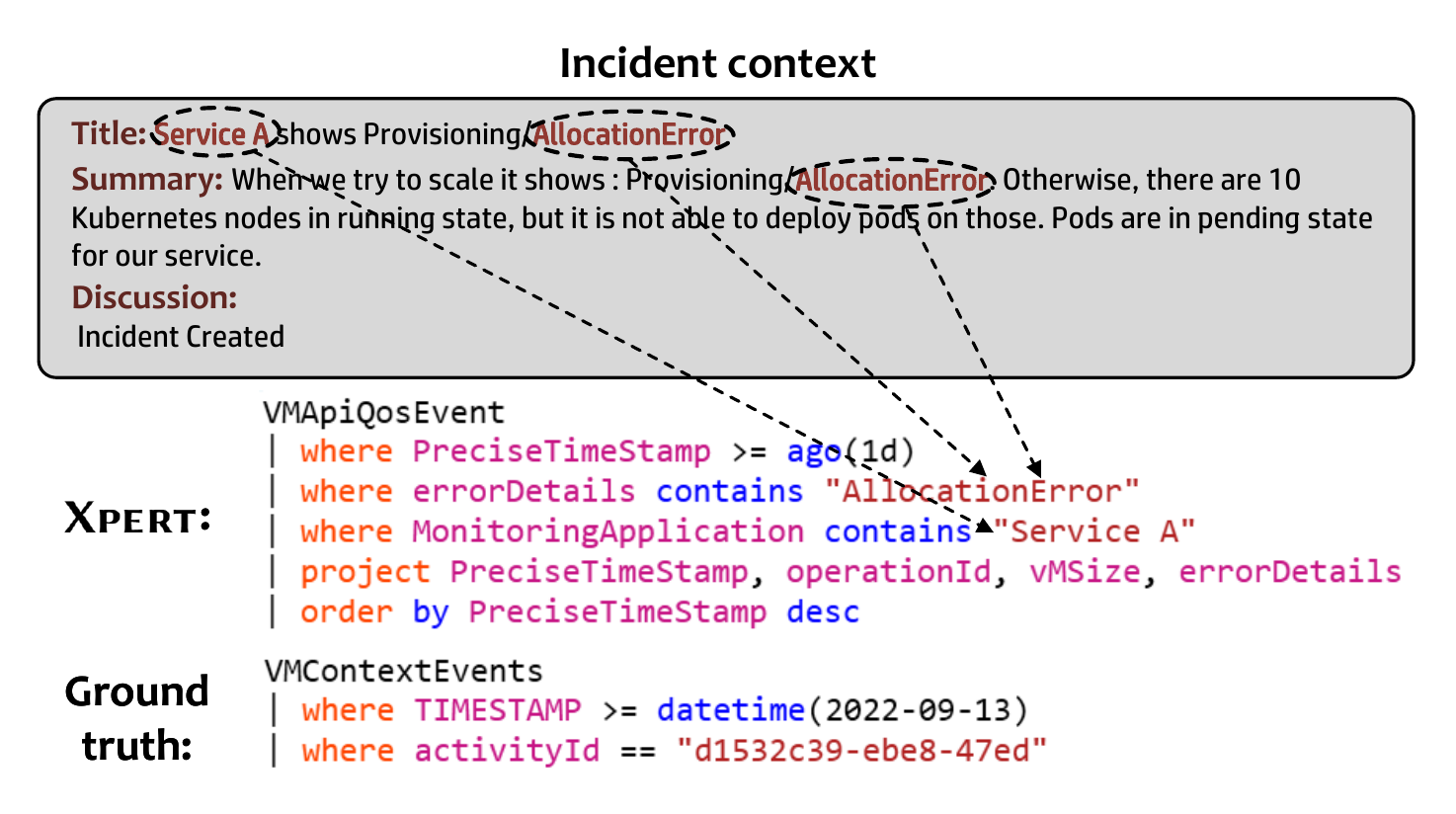}
    \vspace*{-2.5em}
    \caption{A failed case that \name recommends a different KQL query to  the ground truth. \label{fig:badcase}}
    \vspace*{-1.em}
\end{figure}

However, it is essential to acknowledge that \name may not always provide optimal recommendations and may encounter challenges when dealing with incidents that lack crucial information. In Fig.~\ref{fig:badcase}, we illustrate a case where \name fails to recommend the correct query to OCEs. Although the ground truth query is short and simple, crucial information, such as \verb|TIMESTAMP| and \verb|activityId|, is absent from the incident context, rendering it impossible to predict these elements accurately. Upon closer examination of the query recommended by \name, we observe that it still extracts useful information from the incident, such as the error and service details. Additionally, we found that historically, queries similar to the predicted one have occurred frequently in this service. This indicates that the prediction does not come from nowhere,  as \name  encapsulates significant incident information and its predicted query can serve as a valuable reference for OCEs.

\vspace*{-0.5em}
\subsection{Threats to Validity}

\subsubsection{Internal Validity}
Throughout this study, we encountered challenges related to the instability of LLMs provided by OpenAI. The LLM service underwent continuous upgrades with multiple model versions and API changes released. Moreover, hyper-parameter settings, such as temperature and top\_p, may also result in unstable outcomes. To address this issue, we fix the API version and model version and set both temperature and top\_p to zero. Despite these configurations, we observed occasional differences in the LLM response even for the same request. This may be attributed to float precision differences in the underlying hosting machines.
    
The datasets in this study were collected from the incident management platform at \company. Due to the vast volume of incidents over years, we focused on selecting incidents of high severity and top cloud services starting from the year 2022, as they are of greater significance and relevance. While preparing the KQL queries, we extracted the queries from free-text discussions using domain rules. However, this process may introduce inaccuracies or incompleteness in the extracted queries. To mitigate this concern, we meticulously designed and iterated the rules through multiple rounds, and we also removed queries that were found to be invalid.


\subsubsection{External Validity}

While our work focuses on the KQL, our approach is easily applicable to other query languages, such as SQL. The key advantage lies in our method's ability to generalize without requiring model fine-tuning, leveraging the in-context learning capability of LLMs. Additionally, since popular query languages like SQL are extensively encountered during LLM pre-training, we anticipate even better performance when applied to such languages.

Our experiments were conducted on a select set of core services within \company. However, the fundamental concept and workflow of our proposed method can be adapted to other services and scenarios with ease. By incorporating incident-query data from new services, our approach can be readily extended to support additional services. Furthermore, in scenarios with abundant pairwise data, our approach can be adopted to build a versatile pipeline.

\vspace*{-0.5em}
\section{Related Work}
In this section, we examine a selection of notable research studies pertaining to the generation of DSL and the application of language models in the field of software engineering.

\subsection{Domain-Specific Language Generation}
Deep learning techniques have gained significant traction in the DSL generation domain, particularly for languages like SQL \cite{qin2022survey, katsogiannis2023survey}, LaTeX \cite{wang2021translating}, and GraphQL \cite{ni2022knowledge}. These methods aim to make programming languages in specific domains more accessible to non-technical users. Notably, text-to-SQL has received considerable research attention, as SQL is widely used for data querying and insights discovery. In \cite{scholak2021picard}, a fine-tuned T5 model is employed to guide the auto-regressive decoders of language models through incremental parsing. This approach helps minimize the generation of invalid code during the process. Hui \etal propose S\textsuperscript{2}SQL \cite{hui2022s2sql}, which leverages syntactic dependency information from text-to-SQL questions to enhance performance, surpassing a set of pretrained models. Additionally, Liu \etal evaluate the text-to-SQL generation capabilities of ChatGPT \cite{ouyang2022training}, demonstrating its remarkable performance in zero-shot scenarios  \cite{liu2023comprehensive}.


\vspace*{-0.5em}
\subsection{LLMs for Software Engineering}
The advancement of LLMs has greatly contributed to the feasibility and capabilities of these models in the field of software engineering \cite{ahmed2023recommending, chen2023empowering, jin2023assess}. For instance, Ahmed \cite{ahmed2023recommending} employed GPT models to recommend root causes and mitigation steps for incidents, facilitating incident management. Promising results showcased the potential of LLMs in incident management. Similarly, RCACopilot \cite{chen2023empowering} utilized LLMs to improve root cause analysis accuracy by leveraging additional information from troubleshooting guides and chain-of-thought techniques \cite{wei2022chain}. Furthermore, Jin \etal \cite{jin2023assess} employed LLMs to summarize outages in cloud environments. Online performance evaluations demonstrated that LLMs can achieve human-level outage summaries at significantly faster speeds (251.2$\times$). In a departure from the aforementioned practices, \name presents a pioneering framework for automatically recommending DSL queries to support incident management tasks.

\vspace*{-0.5em}
\section{Conclusion}
Incident management plays a critical role in ensuring the smooth functioning of cloud infrastructure, requiring OCEs to execute DSL queries for understanding incidents and support the management processes. This paper conducts a comprehensive empirical study on KQL queries used in a large-scale incident management system at \company, revealing valuable insights into the frequency, complexity, and diversity of KQL queries. Based on these findings, we propose \name, an end-to-end framework that leverages LLMs and ICL to provide customized query recommendations tailored to new incidents, thus empowering the incident management process. To evaluate the quality of the generated queries, we introduce \score, a dedicated evaluation metric that comprehensively assesses query performance from three different perspectives. In offline evaluations using real-world data, \name outperforms other baseline models across various metrics. Furthermore, we have successfully deployed \name in the real production environment of \company, and the piloting results confirm \name's reliability in supporting incident management. To the best of our knowledge, this paper presents the first empirical study of DSL queries for incident management, and \name stands as a pioneering KQL recommendation framework specifically designed to enhance incident management processes.

\section{Data Availability}
The data used in this work originates from the production of \company and contains highly confidential information. We are therefore unable to make the data publicly available due to security concerns.


\clearpage 
\balance
\bibliographystyle{unsrt}
\bibliography{sample-base}


\end{document}